# Detecting and monitoring foodborne illness outbreaks: Twitter communications and the 2015 U.S. *Salmonella* outbreak linked to imported cucumbers

**Abstract**

This research uses Twitter, as a social media device, to track communications related to the 2015 U.S. foodborne illness outbreak linked to *Salmonella* in imported cucumbers from Mexico. The relevant Twitter data are analyzed in light of the timeline of the official announcements made by the Centers for Disease Control and Prevention (CDC). The largest number of registered tweets is associated with the period immediately following the CDC initial announcement and the official release of the first recall of cucumbers. [JEL: I18, Q13, Q18].

**Key words:** foodborne illness, risk communication, Salmonella, Twitter.

**Yuliya V. Bolotova**, Assistant Professor of Agribusiness – *Corresponding Author*
Department of Agricultural Sciences, Clemson University
237 McAdams Hall, Clemson, SC 29634
E-mail: yuliyab@clemson.edu

*Yuliya V. Bolotova is an Assistant Professor of Agribusiness in the Department of Agricultural Sciences at Clemson University. She received a PhD degree in agricultural economics from Purdue University in 2006 and an LLM degree from the University of Chicago Law School in 2010. Her current research program areas include agribusiness economics and strategy, market and price analysis, agricultural and food product marketing and competition law and economics.*

**Jie Lou**, Graduate Student
School of Computing, Clemson University, Clemson, SC 29634
E-mail: jlou@clemson.edu

*Jie Lou is a graduate student in the School of Computing at Clemson University.*

**Ilya Safro**, Assistant Professor
School of Computing, Clemson University
228 McAdams Hall, Clemson, SC 29634
E-mail: isafro@clemson.edu

*Ilya Safro is an Assistant Professor of Computer Science at Clemson University. He received a PhD degree in Applied Mathematics and Computer Science from the Weizmann Institute of Science in 2008. He was a postdoc and Argonne scholar at Mathematics and Computer Science Division of Argonne National Lab. His current research program areas include machine learning, data mining, and algorithms for big data.*

# 1. Introduction

The foodborne diseases cause approximately 48 million people to become ill in the U.S each year (Hofmann et al., 2015). The foodborne illnesses impose enormous economic losses. It was estimated that 9.4 million illnesses imposed an economic burden of $15.5 billion annually (Hofmann et al., 2015). The people awareness of the foodborne illness outbreaks (FIO) may help prevent the spread of disease and decrease the size of the outbreak and economic losses.

During the recent decade, new types of social media (Twitter, Facebook, YouTube, etc.) began affecting people awareness about different events and consequently the choices that they made (Grebitus et al., 2014; Liu and Lopez, 2016). An emerging research (Bitsch et al., 2014; Meyer et al., 2015) suggests to explore how social media can be used in food crisis communication by government authorities.

Our research objective is to use Twitter, as a social media device, to track communications related to the 2015 U.S. FIO linked to *Salmonella* in imported cucumbers from Mexico, the largest in term of the number of affected people outbreak in the country in 2015. As reported by CDC, 907 people infected with the outbreak strains of *Salmonella* Poona were reported from 40 states. We extract and analyze Twitter data in light of the timeline of the official announcements made by CDC during the investigation period[1].

---

[1] CDC is a federal government agency, part of the U.S. Department of Health and Human Services, which is responsible for conducting multistate investigations of FIO.

## 2. Outbreak details

On September 04, 2015 CDC issued the initial public announcement on the FIO linked to *Salmonella* Poona. This bacteria strain was linked to cucumbers imported from Mexico. The affected type of cucumbers is referred to as a "slicer" or "American" cucumber (a dark green in color and typical length is 7 to 10 inches). Retailers sell these cucumbers in a bulk display without any individual packaging or plastic wrapping.

Two voluntarily recalls of cucumbers, which may have been contaminated with *Salmonella*, followed the initial CDC announcement. On September 04 Andrew & Williamson Fresh Produce recalled cucumbers sold under the Limited Edition brand label. On September 11 Custom Produce Sales recalled cucumbers sold under the Fat Boy label. Recalled cucumbers were produced in Baja California, Mexico and distributed in many states within the U.S.

## 3. Twitter data analysis

### 3.1. Data collection and classifying noisy Tweets

Twitter data potentially relevant to the analyzed case were collected in the Clemson University Social Media Listening Center for the overall period of July 2015 to March 2016. This period encompasses two months before the CDC initial announcement and the period including all CDC announcements. While there are millions of tweets that may be registered every day for a particular case, the majority of them are redundant and irrelevant for that case. To eliminate noisy tweets we used the following computational procedure.

At the first stage, we developed and applied keyword-based filters. After a preliminary analysis focusing on a wider range of keywords, we used combinations of the following keywords: Salmonella, Salmonella Poona, Salmonella Tainted, Contaminated Cucumbers, Andrew & Williamson Fresh Produce, Fat Boy Brand and Mexican Cucumbers. While at this stage a lot of irrelevant information was eliminated, the filtered data still contained a significant amount of noise.

At the second stage, our objective was to identify noisy tweets that were not relevant to our study and eliminate them from the dataset. Machine learning offers several effective techniques that can tackle this problem. Examples include clustering, neural networks, decision trees, and Bayesian methods to mention just few of them (Alpaydin, 2014). When choosing a suitable machine learning technique for practical purposes among many that mathematically solve a particular problem, one has to consider the following aspects: running time of the method, quality of the method (including its customization) and availability of software.

We formulate the problem of noise elimination in our dataset as a classification problem, whose solutions belong to a class of supervised machine learning methods. The main objective of the supervised machine learning is to develop a class-labeling prediction function, which will automatically label all elements of a large dataset and newly appearing input elements, given a (typically) small number of already labelled elements (termed as a training set). In the context of our problem, the labels -1 and 1 correspond to "not relevant" and "relevant", respectively. We work with two datasets. The large dataset is a collection of all tweets, and the training set is a small subset of tweets that we manually pre-label. In particular, we identify them as either "relevant" or "not relevant".

We have identified the support vector machine (SVM) technique as the most successful for the purpose of our analysis (Vapnik, 2000). The SVM is among the most well-known optimization-based supervised learning methods, which was originally developed for binary classification problems. The main idea of SVM is to identify a decision boundary with maximum possible margin between the data points of each class. There exist many different classes of SVMs that differ in quality and computational complexity. Training some classes of SVMs (i.e. developing a classifier given a training set) is often a time consuming task when the dataset is large, which requires application of fast heuristics (Razzaghi and Safro, 2015).

In order to apply SVM, we first represent the tweets using traditional information analytic form, namely, the term frequency-inverse document frequency (tf-idf) vector form. We denote the training set of tweets by $D$, with $|D| = N$, and by $f_{j,i}$ the number of times word $j$ appears in tweet $i$. We define the normalized term frequency of word $j$ in tweet $i$ as

$$\text{tf}(j, i) = \frac{1}{2} + \frac{1}{2} \cdot \frac{f_{j,i}}{\max\{f_{k,i} : k \in i\}},$$

and inverse document frequency of term $j$ in $D$ by

$$\text{idf}(j, D) = \log \frac{N}{|\{d \in D : j \in d\}|}.$$

In the tf-idf representation, each tweet $i$ corresponds to a vector $v_i$ in $n$-dimensional space, where the $j$-th entry of $v_i$ is

$$\text{tf-idf}(j, i, D) = \text{tf}(j, i) \cdot \text{idf}(j, D).$$

The tf-idf weight is often used in text mining as a statistical measure to evaluate the importance of a word in a document with respect to the full collection of documents. The importance increases proportionally to the number of times the word appears in the document, but is offset by the frequency of the word in the corpus. A combination of tf-idf and SVM approaches has been demonstrated to be one of the most successful in text classification (Pawar and Gawande, 2012). The tweets in the tf-idf representation form an input of the SVM training process.

We labelled manually the training set $D$, using samples of 100 relevant and 100 irrelevant tweets. The training process results in an SVM classifier function that accepts a tweet in the same tf-idf representation and computes a binary prediction of its class. With this function we process the entire dataset of tweets and eliminate the tweets classified as "not relevant". Since the training set was relatively small to make the computational problem easy enough, no special SVM tools were required. Therefore, we applied an open-source SVM implementation in Python's scikit-learn toolkit.

**3.2. Data analysis**

The number of relevant tweets ("tweets" to be referred further), which have been identified by SVM, is analyzed in light of the timeline of CDC announcements. The frequency of tweets is depicted on Figure 1.

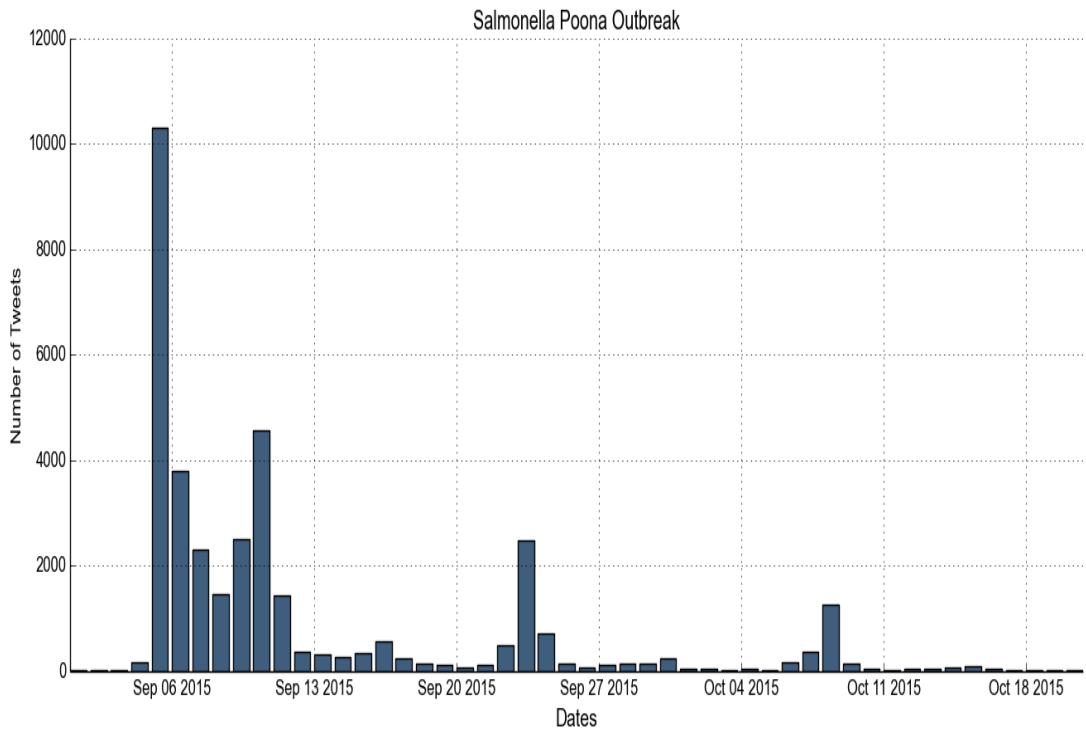

**Figure 1.** 2015 U.S. *Salmonella* Outbreak Linked to Imported Cucumbers: Twitter Data (September 01 - October 20, 2015).

The number of tweets is calculated for each period between the CDC announcements (Table 1).

**Table 1:** 2015 U.S. *Salmonella* Outbreak Linked to Imported Cucumbers:

CDC Announcements, Reported Illnesses and Twitter Data.

| CDC announcement date | Number of registered ill people[*] | Number of tweets[**] |
|---|---|---|
| July 03, 2016[1] | | 442 |
| September 04, 2015[2] | 285 people in 27 states | 18,006 |
| September 09, 2015 | +56 (341 people in 30 states) | 9,425 |
| September 15, 2015 | +77 (418 people in 31 states) | 1,531 |
| September 22, 2015 | +140 (558 people in 33 states) | 4,130 |
| September 29, 2015 | +113 (671 people in 34 states) | 533 |
| October 06, 2015 | +61 (732 people in 35 states) | 2,047 |
| October 14, 2015 | +35 (767 people in 36 states) | 639 |
| November 19, 2015 | +71 (838 people in 38 states) | 1,359 |
| January 26, 2016 | +50 (888 people in 39 states) | 1,582 |
| March 18, 2016[3] | +19 (907 people in 40 states) | 349 |

[*] Cumulative number of registered ill people is in the parentheses.

[**] The number of tweets is calculated for the period following each announcement and prior to the next announcement.

[1] Illness start date (reported later).

[2] Initial CDC announcement.

[3] Final CDC announcement (the number of Tweets is for the period of March 18 to March 31).

A relatively small number of tweets (442) is registered for the period prior to CDC initial announcement. While the illness onset was estimated to be on July 03, 2015, the CDC initial announcement was made on September 04, 2015. By this date, approximately 30% of all registered ill people were identified.

The largest number of tweets corresponds to the period of September 04 to September 28. During this period there were four CDC announcements and two releases on voluntarily recalls of cucumbers from Mexico. The CDC initial announcement and the first recall were issued on September 04. By September 04, 285 people were reported to be sick in 27 states[2]. There were 18,006 tweets registered for the period following CDC initial announcement (September 04) and before the second announcement (September 09). Additional 56 people were reported to be sick.

During the period following CDC second announcement (September 09) and before the third announcement (September 15), the total of 9,425 tweets were registered. The official release of the second voluntarily recall of cucumbers took place on September 11. Additional 77 people were reported to be sick.

During the period of September 15 to October 14 there were five CDC announcements, which were made on a weekly basis. The number of registered tweets for the periods between these announcements included 1,531 following September 15 announcement, 4,130 following September 22 announcement, 533 following September 29 announcement, and 2,047 following October 06 announcement. The largest number of tweets (4,130) followed CDC announcement reporting the largest number of additional ill people registered (140 people).

---

[2] The number of registered sick people is reported in CDC announcements.

The frequency of CDC announcements decreased during the period of October 14, 2015 to March 18, 2016 (CDC final announcement). There were four announcements made during this period. Additional 175 people were reported to be sick. The total of 3,929 relevant Tweets were registered for this period.

## 4. Conclusion

A properly conducted analysis of Twitter data may provide invaluable and timely information for government agencies conducting investigations of FIO. This may potentially allow to reduce the time period between the beginning of the outbreak and the initial public announcement and to result in a quicker communication of associated risks.

Using new types of social media (Twitter, Facebook, etc.) for detecting and monitoring FIO is an emerging and promising research area that offers important insights and implications for business and policy decision making process. A crucial component in a proper evaluation of social media data is developing and evaluating advanced techniques that can discover emergent information in real time (Avudaiappan et al., 2016). These techniques would allow obtaining information that government agencies may use, while implementing preventive measures, investigating outbreaks and communicating associated food safety risks to public.